\documentclass[twocolumn,10pt]{autart}
\usepackage{mathrsfs}
\usepackage{graphicx}
\usepackage{amsmath}
\usepackage{enumerate}
\usepackage{graphicx}
\usepackage{tikz}
\usepackage{apacite}
\usepackage{natbib}
\usepackage{amsfonts}
\usepackage{bm}
\usepackage{amssymb}
\usepackage{subfigure}
\usepackage{hyperref}
\usepackage{float}
\hypersetup{
	colorlinks = true,
	citecolor = cyan,
}
\usepackage{wrapfig}
\newenvironment{proof}{{\indent \indent \bf Proof.}}{\hfill $\qed$\par}
\newtheorem{remark}{Remark}

\newtheorem{assumption}{Assumption}
\newtheorem{example}{Example}
\newtheorem{lemma}{Lemma}
\newtheorem{theorem}{Theorem}

\def\ban{\begin{eqnarray*}}
\def\ean{\end{eqnarray*}}
\def\bna{\begin{eqnarray}}
\def\ena{\end{eqnarray}}

\begin{document}
	
	\begin{frontmatter}
		
		\title{Distributed Sparse Identification for Stochastic Dynamic Systems under Cooperative Non-Persistent Excitation  Condition\thanksref{footnoteinfo}}

		\thanks[footnoteinfo]{This work was supported by the National Key R\&D Program of China under Grant 2018YFA0703800, the Strategic Priority Research Program of Chinese Academy of Sciences under Grant No. XDA27000000, Natural Science Foundation of China under Grant 11688101 and U21B6001, and National Science Foundation of Shandong Province (ZR2020ZD26). Corresponding author: Zhixin Liu.}
		\author[AMSS,CAS]{Die Gan}\ead{gandie@amss.ac.cn},
		\author[AMSS,CAS]{Zhixin Liu}\ead{lzx@amss.ac.cn}
		
		\address[AMSS]{Key Laboratory of Systems and Control, Institute of Systems Science, Academy of Mathematics and Systems Science,\\
			Chinese Academy of Sciences, Beijing 100190, P. R. China.}
		\address[CAS]{School of Mathematical Sciences, University of Chinese Academy of Sciences, Beijing 100049, P. R. China.}
		
		
\begin{abstract} This paper considers the distributed sparse identification problem over wireless sensor networks such that all sensors cooperatively estimate the unknown sparse parameter vector of stochastic dynamic systems by using the local information from neighbors. A distributed sparse least squares algorithm is proposed by minimizing a local information criterion formulated as a linear combination of accumulative local estimation error and $L_1$-regularization term.
The upper bounds of the estimation error and the regret of the adaptive predictor of the proposed algorithm are presented. Furthermore, by designing a suitable adaptive weighting coefficient based on the local observation data,  the set convergence of zero elements with a finite number of observations is obtained under a cooperative non-persistent excitation condition. It is shown that the proposed distributed algorithm can work well in a cooperative way even though none of the individual sensors can fulfill the estimation task. Our theoretical results are obtained
without relying on the independency assumptions of regression signals that have been commonly used in the existing literature. Thus, our results are expected to be applied to stochastic feedback systems. Finally, the numerical simulations are provided to demonstrate the effectiveness of our theoretical results.
		\end{abstract}
		
	\end{frontmatter}
	
	\section{Introduction}
In recent years, wireless sensor networks (WSNs) have attracted increasing research attention
because of their wide application in engineering systems including smart grids, biomedical health monitoring, target tracking and surveillance
\citep{article:Sayed2013,article:Yick2008}.
Distributed observation and data analysis are ubiquitous in WSNs, where sensors are interconnected to acquire and
process the local information from neighbors to finish a common task.
Due to various uncertainties in practical systems, the distributed identification problem over WSNs becomes one of the important topics where all
the sensors collaboratively estimate an unknown parameter vector of interest by using
local noisy measurements. Unlike the centralized method with a fusion center, the distributed scheme has the advantages of flexibility, robustness to node or link failures as well as reducing communication load and calculation pressure.  Consequently, the theoretical analysis of distributed estimation or filtering algorithms based on several typical distributed strategies
such as the incremental, the diffusion and the consensus strategies have been provided \citep{Abdolee2016, Jian2017, Stefano2020,FULLY}.

In practical scenarios, there exist a
large number of sparse systems \citep{Bazerque2010,Vinga2021} where many elements in the parameter vector do not contribute or contribute marginally to the systems ( i.e., these elements are zero or
near-zero).  How to infer the zero elements and identify the nonzero elements in the unknown parameter vector is an important issue in the investigation of sparse systems.
Considerable progress has been made on the identification of zero and nonzero elements in an unknown sparse parameter vector \citep{Peng2006, Chiuso2014, Eksioglu2013}, which allows us to obtain a more reliable prediction model.
One direction for the estimation of sparse signals is based on the compressed sensing (CS) theory \citep{can2,com7}, and some estimation algorithms using CS are proposed (cf., \cite{com12,com13}) in which \emph{a priori} knowledge about the sparsity of the unknown parameter and the regression vectors are required. Another direction is the sparse optimization based on the regularization framework  where the objective function is formulated as a combination of the prediction error with a penalty term. The well-known LASSO (the least absolute shrinkage and selection operator) is one of the classical algorithms to obtain the sparse signals \citep{Tibshirani1996}, and its variants and adaptive LASSO \citep{Zou2006} are also studied. For the stochastic dynamic systems with a single sensor, the adaptive sparse estimation or filtering algorithms are studied by combing the recursive least squares (LS) and  least mean squares (LMS) with regularization term \citep{zhao, Chen_Gu2009}.

With the development of sensor networks, some distributed adaptive sparse estimation algorithms have been proposed, and  the corresponding stability and  convergence analysis are also investigated
under some signal conditions. For example,  \cite{Lorenzo2013} provided the convergence and mean-square performance analysis for the distributed LMS
algorithm regularized by convex penalties where the assumption of independent regressors is required.  \cite{Huang2015}  presented theoretical analysis on the mean and mean-square
performance of the distributed sparse total LS algorithm under the condition that the input  signals are independent and identically distributed (i.i.d.).  \cite{Shiri2018} analyzed the mean stability of distributed quasi-sparse affine projection algorithm with  independent  regression vectors. \cite{WeiHuang2020} analyzed the mean stability of the sparse diffusion LMS algorithm for two regularization terms with independent regression vectors.  However, for the typical models such as ARMAX (autoregressive moving-average with exogenous input) model and Hammerstein system, the regressors are often generated by the past input and output signals, so it is hard for them to satisfy the aforementioned independency assumptions.

In order to relax the independency assumption of the regressors, some attempts are made for the
distributed adaptive estimation or filtering algorithms. For the unknown time-invariant parameter vector, \cite{Gan2019} proposed a distributed stochastic gradient algorithm, and established the strong consistency of the proposed algorithm under a cooperative excitation condition.
 \cite{wc2} studied the convergence of the diffusion LS algorithm. For the time-varying parameter vector, \cite{XIE2018} provided a cooperative information condition to guarantee the stability of the consensus-based LMS adaptive filters. Moreover, \cite{Gan2021} introduced the collective random observability
condition and provided the stability analysis of the distributed Kalman filter algorithm.
Nevertheless, these asymptotical results are established  as the number of the observation data obtained by sensors tends to infinity, which may not be suitable for  the sparse identification problem with limited observation data.

Inspired by  \cite{zhao} where a sparse identification algorithm for a single sensor case is put forward to infer the set of zero elements with finite observations, we develop a distributed adaptive sparse LS algorithm over sensor networks such that all sensors can cooperatively identify the unknown parameter vector and infer the zero elements with a finite number of observations.
The main contributions can be summarized as follows:
\begin{itemize}
\item We first introduce a local information criterion for each sensor which is formulated as a linear combination of local estimation errors  with $L_1$-regularization term.  By minimizing this criterion, a distributed adaptive sparse identification algorithm is proposed. The upper bounds of the estimation error
and the accumulative regret of the adaptive predictor are established, which can be degenerated to the results of the classical distributed LS  algorithm \citep{wc2} when the weighting coefficients are equal to zero.
\item
Then, we introduce a cooperative non-persistent excitation condition on the regressors, under which  the distributed
sparse LS algorithm can cooperatively identify the set of zero elements with finite observations by properly choosing the weighting coefficients. 
	We remark that the key difference between the proposed algorithm and those in distributed  sparse optimization framework (e.g., \cite{Lorenzo2013})  lies in that the weighting coefficients are generated from the local observation sequences.
 The cooperative excitation condition is much weaker than the widely used persistent excitations  (cf., \cite{Chen2014, Zhang2021,sheng2015}) and the regularity condition \citep{Zou2006}.


\item Different from most existing results on the distributed sparse algorithms, our theoretical results are obtained
without relying on the independency assumptions of regression signals, which makes it possible
for applications to the stochastic feedback systems. We also reveal that the whole sensor network can cooperatively accomplish the estimation task,  even if any individual sensor can not due to lack of necessary information \citep{zhao}.
\end{itemize}

The remainder of this paper is organized as follows. In
Section \ref{problem_formulation}, we give the problem formulation of this paper; Section \ref{main_results} presents the main results of the paper including the parameter convergence of the algorithm, the regret analysis, and the set convergence of the algorithm; the proofs of the main results
are given in Section \ref{proofs_of}. A simulation example is provided in Section \ref{simulaiton}. Finally, we conclude the
paper with some remarks in Section \ref{concluding}.

\section{Problem formulation}\label{problem_formulation}
\subsection{Basic notations}
In this paper,  for an $m$-dimensional vector $\bm x$, its $L_p$-norm is defined as $\|\bm x\|_{p}=(\sum^m_{j=1}|\bm x(j)|^p)^{1/p}$  ($1\leq p<\infty$) , where $\bm x(j)$ denotes the $j$-th element of  $\bm x$.  For $p = 1$, $\|\bm x\|_{1}$ is the sum of absolute values
of all the elements in $\bm x$; and for $p = 2$, $\|\bm x\|_{2}$ is the Euclidean norm, we simply write $\|\cdot\|_{2}$ as $\|\cdot\|$.
 For an $m\times m$-dimensional real matrix $\bm A$, we use $\lambda_{\max}(\cdot)$ and $\lambda_{\min}(\cdot)$ to denote the largest and smallest eigenvalues of the matrix.
  $\|\bm A\|$ denotes the Euclidean norm, i.e., $\|\bm A\|=(\lambda_{max}(\bm A\bm A^T))^{\frac{1}{2}}$ where the notation $T$ denotes the transpose operator; $\|\bm A\|_F$ denotes the Frobenius norm, i.e., $\|\bm A\|_F=(tr(\bm A^T\bm A))^{\frac{1}{2}}$, where the notation $tr(\cdot)$ denotes the trace of the corresponding matrix.  	We use $col(\cdot,\cdots,\cdot)$ to denote a vector stacked by the specified vectors, and $diag(\cdot,\cdots,\cdot)$ to denote a block matrix formed in a diagonal manner of the corresponding vectors or matrices.
For a symmetric matrix $\bm A$, if all eigenvalues of $\bm A$ are positive (or nonnegative), then it is a positive definite (semipositive) matrix, and we denote it as  $\bm A>0~(\geq 0)$.
If all elements of a matrix $\bm A=\{a_{ij}\}\in\mathbb{R}^{n\times n}$ are nonnegative, then it is a nonnegative matrix, and furthermore if  $\sum^n_{j=1}a_{ij}=1$ holds for all $i\in\{1,\cdots,n\}$, then it is called a stochastic matrix.

For any two positive  scalar
sequences $\{a_k\}$ and $\{b_k\}$,
by $ a_k = O(b_k)$ we mean that there exists a constant $C > 0$ independent of $k$ such that
$a_k \leq  C b_k$ holds for all $k \geq  0$, and by $a_k=o(b_k)$ we mean that $\lim_{k\rightarrow \infty}a_k/b_k=0$.
 For a convex function $f(x)$, we use
$\partial f:x\rightarrow\partial f(x)$  to denote the subdifferential of $f$, which is a convex set. For example,
\ban \partial |x|=\left\{
                    \begin{array}{ll}
                      1, & \hbox{if}\ x>0; \\
                      -1, & \hbox{if}\ x<0; \\
                     $[-1,1]$, & \hbox{if} \ x=0,
                    \end{array}
                  \right.  .
 \ean A necessary and sufficient condition that a given point $x$ belongs to the minimum set of $f$ is  $0\in\partial f(x)$ (see \cite{Rockafellar1972}). We also need to introduce the sign function $\rm{sgn}$$(x)$ defined as   $\rm{sgn}$$(x)=1$ if $x\geq0$ and  $\rm{sgn}$$(x)=-1$ if  $x<0$.

\subsection{Graph theory}
	 We consider a sensor network with $n$ sensors. The communication between sensors
 are usually modeled as an undirected weighted graph $\mathcal{G}=(\mathcal{V},\mathcal{E},$
$\mathcal{A})$,  where $\mathcal{V}=\{1,2,3,\cdots, n\}$ is the set of sensors (or nodes), $\mathcal{E}\subseteq\mathcal{V} \times\mathcal{V}$ is the edge set, and $\mathcal{A}=\{a_{ij}\}\in\mathbb{R}^{n\times n}$ is the weighted adjacency matrix. The elements of the
adjacency
matrix $\mathcal{A}$ satisfy  $a_{ij}>0$ if $(i, j)\in \mathcal{E}$ and $a_{ij}=0$ otherwise.
Here we assume that the matrix
$\mathcal{A}$ is a symmetric and stochastic matrix.  For the sensor $i$, the set of its neighbors is denoted as $N_i=\{j\in \mathcal{V}|(i,j)\in\mathcal{E} \}$, and the sensor $i$ belongs to $N_i$.
The sensor $i$ can communicate  information with its neighboring sensors. A path of length $\ell$ is a sequence of nodes $\{i_1,..., i_{\ell}, i_{\ell+1}\}$ such that $(i_h, i_{h+1})\in \mathcal{E}$ with $1\leq h\leq \ell$. The graph $\mathcal{G}$ is called connected if there is a path between any two sensors. The diameter $D_\mathcal{G}$ of the graph $\mathcal{G}$ is defined as the maximum shortest path length between any two sensors.

\subsection{Observation model}
	In this paper, we consider the parameter identification problem in a network consisting of $n$ sensors labeled $1,\cdots,n$.
Assume that the data $\{y_{t,i}, \bm\varphi_{t,i}, t=1,2,\cdots\}$ collected by the sensor $i$ obeys the following discrete-time stochastic regression model,
\bna
y_{t+1,i}=\bm \varphi_{t,i}^T\bm \theta+w_{t+1,i}, ~t=0, 1, 2, \cdots, \label{model}
\ena
where $y_{t,i}$ is the scalar observation or output of the sensor $i$ at time $t$, $\bm\varphi_{t,i}$ is the
$m$-dimensional stochastic regression vector which may be the function of current and past
inputs and outputs, $\bm\theta\in \mathbb{R}^m$ is an unknown $m$-dimensional parameter to be estimated, and
$\{w_{t,i}\}$ is the noise sequence. The above model (\ref{model}) includes many parameterized systems,
such as ARX system and Hammerstein system.
We further denote the parameter vector $\bm\theta$ and the index set of its zero elements by
\begin{equation}
\begin{split}
\bm\theta&\triangleq(\bm\theta(1),\cdots,\bm\theta(m))^T,\\
H^*&\triangleq\{l\in \{ 1,\cdots,m\}|\bm\theta(l)=0\}.\label{sparse5}
\end{split}
\end{equation}

Our problem is to design a distributed adaptive estimation algorithm such that all sensors cooperatively infer the set $H^*$ in a finite number of steps and identify the unknown parameter $\bm\theta$  by using stochastic regression vectors and the
observation signals  from its neighbors,
i.e., $\{\bm \varphi_{k,j}, y_{k+1,j}\}^t_{k=1}~(j\in N_{i})$.

	\section{The main results}\label{main_results}

\subsection{Parameter convergence}

Before designing the algorithm to cooperatively estimate the unknown parameter vector and infer the set $H^*$, we first introduce the following classical distributed least squares algorithm to estimate the unknown parameter $\bm \theta$ in (\ref{sparse5}), i.e.,
\bna
\bm\theta_{t+1,i}=\bm P_{t+1,i}\left(\sum^n_{j=1}
\sum^t_{k=0}a^{(t+1-k)}_{ij}\bm\varphi_{k,j}y_{k+1,j}\right),\label{theta}
\ena
where $
\bm P_{t+1,i}=\left(\sum^n_{j=1}\sum^t_{k=0}a^{(t+1-k)}_{ij}\bm\varphi_{k,j}\bm\varphi^T_{k,j}\right)^{-1}$ and $a^{(t+1-k)}_{ij}$ is the $i$-th row, $j$-th column entry of the matrix $\mathcal{A}^{t+1-k}$.   It is clear that the matrix $\bm P_{t+1,i}$
can  be equivalently written as the following recursive form,
\begin{gather}
\bm P^{-1}_{t+1,i}=\sum_{j\in N_i}a_{ij}({{\bm P}}^{-1}_{t,j}+\bm\varphi_{t,j}\bm\varphi^T_{t,j}).\label{P_inverse}
\end{gather}
Thus, the algorithm (\ref{theta}) can also have the following recursive expression,
\begin{gather}
\bm \theta_{t+1,i}=\bm P_{t+1,i}\sum_{j\in N_i}a_{ij}(\bm P^{-1}_{t,j}\bm \theta_{t,j}+\bm \varphi_{t,j}y_{t+1,j}).
\label{theta1}
\end{gather}
Note that in the above derivation, we assume that the matrix $\sum^n_{j=1}\sum^t_{k=0}a^{(t+1-k)}_{ij}\bm\varphi_{k,j}\bm\varphi^T_{k,j}$ is invertible which is usually not satisfied for small $t$. To solve this problem, we take the initial  matrix $\bm P_{0,i}$ to be positive definite. By (\ref{P_inverse}), we have
\bna
\bm P^{-1}_{t+1,i}=\sum^n_{j=1}\sum^t_{k=0}a^{(t+1-k)}_{ij}\bm\varphi_{k,j}\bm\varphi^T_{k,j}
+\sum^n_{j=1}a^{(t+1)}_{ij}\bm P^{-1}_{0,j}.\label{sparse16}
\ena
This modification will not affect the analysis of the asymptotic properties of the estimate of the distributed least squares algorithm. 

In fact, the algorithm (\ref{theta1}) can be obtained by minimizing  the following  linear combination of the estimation error $\sigma_{t+1,i}(\bm\beta)$ between the observation signals and the prediction of the local neighbors,
\bna
\sigma_{t+1,i}(\bm\beta)&=&\sum_{j\in N_i}a_{ij}\bigg(\sigma_{t,j}(\bm\beta)
+[y_{t+1,j}-{\bm\beta}^T\bm\varphi_{t,j}]^2\bigg),\label{least}
\ena
with $\sigma_{0,i}(\bm\beta)=0$. That is, $\bm\theta_{t+1,i}\triangleq \arg\min_{\bm\beta}\sigma_{t+1,i}(\bm\beta)$.

  Set
\ban
\bm e_{t+1}(\bm\beta)&=&col\{(y_{t+1,1}-{\bm\beta}^T\bm\varphi_{t,1})^2,
...,(y_{t+1,n}-{\bm\beta}^T\bm\varphi_{t,n})^2\},\\
\bm{\sigma}_{t}(\bm\beta)&=&col\{\sigma_{t,1}(\bm\beta),...,\sigma_{t,n}(\bm\beta)\}.
\ean
Hence by (\ref{least}), we have
\ban
\bm{\sigma}_{t+1}(\bm\beta)&=&\mathcal{A}\bm{\sigma}_{t}(\bm\beta)+\mathcal{A}\bm e_{t+1}(\bm\beta)\\
&=&\mathcal{A}^2\bm{\sigma}_{t-1}(\bm\beta)+\mathcal{A}^2\bm e_{t}(\bm\beta)+\mathcal{A}\bm e_{t+1}(\bm\beta)\\
&=&\sum^t_{k=0}\mathcal{A}^{t+1-k}\bm e_{k+1}(\bm\beta),
\ean
which implies that
\bna
\sigma_{t+1,i}(\bm\beta)=\sum^n_{j=1}\sum^{t}_{k=0}a^{(t+1-k)}_{ij}
[y_{k+1,j}-{\bm\beta}^T\bm\varphi_{k,j}]^2. \label{least2}
\ena

It is shown by  \cite{wc2} that the distributed least squares algorithm (\ref{theta1}) can generate a consistent estimate for the unknown parameter when the number of data tends to infinity. However, for the sparse unknown parameter vectors (i.e., there are many zero elements in $\bm\theta$), it is hard to infer the zero elements in a finite step due to  the limitation of observations in practice. In order to solve this issue, we introduce the following local information
criterion with $L_1$-regularization to identify the unknown sparse parameters and infer the set $H^*$,
\bna
J_{t+1,i}(\bm\beta)=\sigma_{t+1,i}(\bm\beta)+\alpha_{t+1,i}\|\bm\beta\|_1,\label{penalty}
\ena
where $\|\cdot\|_1$ is the $L_1$-norm, $\alpha_{t+1,i}$ is the weighting coefficient chosen to satisfy $\alpha_{t+1,i}=o(\lambda_{\min}(\bm P^{-1}_{t+1,i}))$, and  $\sigma_{t+1,i}(\bm\beta)$ is recursively defined by (\ref{least}).
For the sensor $i$,  we can obtain the following distributed sparse LS algorithm to estimate the unknown parameter $\bm \theta$ by minimizing $ J_{t+1,i}(\bm\beta)$, i.e.,
\bna
\bm\beta_{t+1,i}= \arg\min_{\bm\beta} J_{t+1,i}(\bm\beta).\label{beta}
\ena

\begin{remark} For the sensor $i$,
the coefficients $\alpha_{t+1,i}$  in (\ref{penalty}) can be dynamically adjusted by using the local observation sequence
$\{\bm\varphi_{k,j}, y_{k+1,j}, j\in N_{i}\}^t_{k=1}$, which makes (\ref{penalty}) be the adaptive LASSO (cf., \cite{Zou2006}).  We show that by properly choosing the coefficient $\alpha_{t+1,i}$, we can identify the set of the zero elements in the unknown sparse parameter vector $\bm\theta$ with a finite number of observations (see Theorem \ref{theorem2}).
\end{remark}

In the following, we will first investigate the upper bound of the estimation error generated by (\ref{beta}), which provides the basis for the set convergence of zero elements. For this purpose, we need to introduce the following assumptions on  the network topology and the observation noise.
\begin{assumption}\label{a1}
The communication graph $\mathcal{G}$ is connected.
\end{assumption}

\begin{remark} For the weighted adjacency matrix $\mathcal{A}$ of the graph $\mathcal{G}$, we denote $\mathcal{A}^l\triangleq(a_{ij}^{(l)})$ with $l\geq 1$.  By the theory of product of stochastic matrices, we see that under Assumption \ref{a1}, $\mathcal{A}^l$ is a positive matrix for $l\geq D_{\mathcal{G}}$, i.e., for any $i$ and $j$, $a_{ij}^{(l)}>0$.
\end{remark}

\begin{assumption}\label{a2}
For any $i\in\{1,\cdots,n\}$, the noise sequence $\{w_{k,i},\mathscr{F}_k\}$ is a martingale difference, and
there exists a constant $\delta > 2$
such that~
$$
\sup_{k\geq 0}E[|w_{k+1,i}|^{\delta}|\mathscr{F}_k]<\infty, {~~\rm a.s.},
$$
where
$\mathscr{F}_t=\sigma\{\bm \varphi_{k,i}, w_{k,i}, k\leq t, i=1,\cdots,n\}$  is a sequence of nondecreasing $\sigma$-algebras and $E[\cdot|\cdot]$ denotes  the conditional expectation operator.
\end{assumption}

  We can verify that the i.i.d. zero-mean bounded or  Gaussian noise $\{w_{k,i}\}$ which 
are independent of the regressors can satisfy Assumption \ref{a2}.

Assume that there are $d$ nonzero elements in the unknown parameter vector $\bm\theta$. Without loss of generality, we assume
$\bm\theta=(\bm\theta(1),\cdots,\bm\theta(d),\bm\theta(d+1),\cdots,\bm\theta(m))^T$ with
$\bm\theta(l)\neq 0, l=1,\cdots,d,$ and  $\bm\theta(j)=0, j=d+1,\cdots,m.$ For the estimate $\bm\beta_{t+1,i}$  obtained by the distributed sparse LS algorithm (\ref{beta}), we denote the estimate error as
\begin{gather}
\widetilde{\bm\beta}_{t+1,i}=\bm\beta_{t+1,i}-\bm\theta. \label{sparse15}
\end{gather}
Then we have the following result concerning the upper bound of the estimation error $\widetilde{\bm\beta}_{t,i}$.

\begin{theorem}\label{theorem1}
Let $\bm P^{-1}_{t+1,i}$ be generated by (\ref{P_inverse}) with arbitrarily initial matrix $\bm P_{0,i}>0$.
 Then under Assumptions \ref{a1} and \ref{a2},
we have for all $i\in\{1,\cdots,n\}$
\ban
\|\widetilde{\bm\beta}_{t+1,i}\|=O\left(\frac{\alpha_{t+1,i}}
{\lambda_{\min}(\bm P^{-1}_{t+1,i})}+\sqrt{\frac{\log r_t}{\lambda_{\min}(\bm P^{-1}_{t+1,i})}}\right),  {\rm a.s.}
\ean
where $r_t=\max\limits_{1\leq i\leq n}\lambda_{\max}\{\bm P^{-1}_{0,i}\}+\sum^n_{i=1}\sum^t_{k=0}\|\bm\varphi_{k,i}\|^2$.
\end{theorem}
The proof of Theorem \ref{theorem1} is provided in Subsection \ref{proof:theorem1}.
\begin{remark}\label{remark_two}
By (\ref{sparse16}), we have
for $t\geq D_{\mathcal{G}}$,
\begin{gather}
\lambda_{\min}(\bm P^{-1}_{t+1,i})\geq a_{\min}\lambda^{n,t}_{\min},\label{sparse28}
\end{gather}
where $a_{\min}\triangleq \min_{i,j\in\mathcal{V}}a^{(D_{\mathcal{G}})}_{ij}>0$ and
\ban
\lambda^{n,t}_{\min}= \lambda_{\min}\left\{\sum^n_{j=1}\bm P^{-1}_{0,j}+\sum^n_{j=1}\sum ^{t-D_{\mathcal{G}+1}}_{k=0}\bm \varphi_{k,j}{\bm \varphi}^T_{k,j}\right\}.\ean  From Theorem \ref{theorem1},
if the coefficient  $\alpha_{t+1,i}$ is chosen to satisfy  $\alpha_{t+1,i}=o(\lambda_{\min}(\bm P^{-1}_{t+1,i}))$ and the regression vectors satisfy the weakest possible cooperative excitation condition $\log r_t=o(\lambda^{n,t}_{\min})$ (cf., \cite{wc2}), then the almost sure convergence of the distributed sparse LS algorithm can be obtained, i.e., $\bm\beta_{t+1,i}\xrightarrow[{t\rightarrow\infty}] {} \bm\theta$.

\end{remark}


\subsection{Analysis of the regret}
Regret is one of the key metrics for evaluating the performance of the online learning algorithms \citep{ieee5, ieee6}.
For each sensor $i\in\{1,\cdots,n\}$, we construct an adaptive predictor $\hat{y}_{t+1,i}$ by using the estimate $\bm \beta_{t,i}$ defined in (\ref{beta}) at the time instant $t$,
$$\hat{y}_{t+1,i}=\bm \varphi^T_{t,i}\bm \beta_{t,i}.$$ The prediction error can be described by the following loss function $\rho_{t+1,i}(\bm \beta_{t,i})$, i.e.,
\ban
\rho_{t+1,i}(\bm \beta_{t,i})&=& E\left[(y_{t+1,i}-\hat{y}_{t+1,i})^2|\mathscr{F}_k\right]\\
&=& E\left[(y_{t+1,i}-\bm \varphi^T_{t,i}\bm \beta_{t,i})^2|\mathscr{F}_t\right].
\ean
 Then the cumulative  regret over the whole network is defined as
\ban
R_{t}=\sum^n_{i=1}\sum^t_{k=0}\rho_{k+1,i}(\bm \beta_{k,i})-\min_{\bm \zeta\in \mathbb{R}^m}\sum^n_{i=1}\sum^t_{k=0}\rho_{k+1,i}(\bm \zeta).
\ean
The regret defined above reflects the difference between the cumulative loss $\rho_{k+1,i} (\bm \beta_{k,i})$ when the unknown parameter is estimated by  (\ref{beta}) and the optimal static value of the cumulative loss function $\rho_{k+1,i} (\cdot)$.  Due to existence of the noise, it is generally  desired that the average regret  $R_{t}/nt$ is small or even goes to zero as  $t\rightarrow\infty$.
	
In the following, we analyze the asymptotic property of the regret $R_{t}$ over the sensor network.
By Assumption \ref{a2}  and the fact $\bm \varphi^T_{k,i}\widetilde{\bm \beta}_{k,i}\in\mathscr{F}_k$, we have
\bna
R_t&=&\sum^n_{i=1}\sum^t_{k=0}E((y_{k+1,i}-\hat{y}_{k+1,i})^2|\mathscr{F}_k)\nonumber\\
&&-\min_{\bm \zeta\in\mathbb{R}^m}\sum^n_{i=1}\sum^t_{k=0}E((y_{k+1,i}-\bm \varphi^T_{k+1,i}\bm \zeta)^2|\mathscr{F}_k)\nonumber\\
&=&\sum^n_{i=1}\sum^t_{k=0}E(\bm \varphi^T_{k,i}\widetilde{\bm \beta}_{k,i}+w_{k+1,i})^2|\mathscr{F}_k)\nonumber\\
&&-\min_{\bm \zeta\in\mathbb{R}^m}\sum^n_{i=1}\sum^t_{k=0}E((\bm \varphi^T_{k,i}({\bm \theta}-\bm \zeta)+w_{k+1,i})^2|\mathscr{F}_k)\nonumber\\
&=&\sum^n_{i=1}\sum^t_{k=0}(\bm \varphi^T_{k,i}\widetilde{\bm \beta}_{k,i})^2.\label{ee1}
\ena

\begin{theorem}\label{theorem4}
Under Assumption \ref{a2}, if ~$\bm\Phi^T_{t}\bm P_t\bm\Phi_{t}=O(1)$, and $\alpha_{t,i}=O\Big(\sqrt{\lambda_{\min}(\bm P^{-1}_{t,i})}\Big)$, we have
\ban
R_t=O(\log r_t), ~~{\rm a.s.}
\ean
where ~  $\bm \Phi_t\triangleq diag\{\bm\varphi_{t,1},...\bm\varphi_{t,n}\}$,  $\bm P_t\triangleq diag\{\bm P_{t,1},...,\bm P_{t,n}\},$ and
 $r_t$ is defined in Theorem \ref{theorem1}.
\end{theorem}
The proof of Theorem \ref{theorem4} is given in Subsection \ref{proof:theorem2}.

\begin{remark}
We know that for the bounded regressors $\bm\varphi_{t,i}$, $r_t$ will be
of the order $O(t)$. Consequently, by  Theorem \ref{theorem4},  the upper bound of the regret $R_t$ over the sensor network is sublinear with respect to $nt$, i.e.,  $R_{t}/nt=O({\log t}/{t})\rightarrow 0$ as $t\rightarrow \infty$. The analysis of the regret does not require
any excitation condition on the regression signals.
 Theorem \ref{theorem1}  and Theorem \ref{theorem4} can be degenerated to the   results of the classical distributed LS  algorithm in  \cite{wc2}  when  $\alpha_{t+1,i}$ is equal to zero.
\end{remark}
\subsection{Set convergence}

In the last two subsections, we have obtained the asymptotic results concerning the parameter convergence and the regret analysis. Inspired by \cite{zhao}, we propose the following distributed sparse adaptive algorithm (Algorithm \ref{algorithm2}) to
identify the set of zero elements with a finite number of observations by choosing  $\alpha_{t,i}$ adaptively.
\begin{algorithm}[htb] 
{\caption{ }\label{algorithm2}
\textbf{Step 1:}
Based on $\{\bm \varphi_{k,j}, y_{k+1,j}\}^t_{k=1}~(j\in N_{i})$, begin with an initial vector $\bm\theta_{0,i}$ and  an initial matrix $\bm P_{0,i}>0$,
compute the matrix $\bm P^{-1}_{t+1,i}$ defined by (\ref{P_inverse}) and the local estimate $\bm\theta_{t+1,i}$ of $\bm\theta$  by (\ref{theta1}), and further define
\bna
&&\hat{\bm \theta}_{t+1,i}(l)\nonumber\\
&\triangleq& \bm\theta_{t+1,i}(l)+{\rm{sgn}}(\bm\theta_{t+1,i}(l) )\sqrt{\frac{\log(\lambda_{\max}(\bm P^{-1}_{t+1,i}))}{\lambda_{\min}(\bm P^{-1}_{t+1,i})}},\label{sparse6}
\ena

\textbf{Step 2:} Choose a positive sequence $\{\alpha_{k,i}\}^{t+1}_{k=1}$ satisfying
\bna
&&\alpha_{k,i}=o(\lambda_{\min}(\bm P^{-1}_{k,i})),\nonumber\\
&&\lambda_{\max}(\bm P^{-1}_{k,i})\sqrt{\frac{\log(\lambda_{\max}(\bm P^{-1}_{k,i}))}{\lambda_{\min}(\bm P^{-1}_{k,i})}}=o (\alpha_{k,i}).\label{sparse27}
\ena

\textbf{Step 3:} Optimize the convex
objective local function,
\bna
 \bar J_{t+1,i}(\bm\xi)=\sigma_{t+1,i}(\bm\xi)+\alpha_{t+1,i}\sum^m_{l=1}\frac{1}{|\hat{\bm \theta}_{t+1,i}(l)|}|\bm \xi(l)| \label{sparse4}
\ena}
with $\sigma_{t+1,i}(\bm\xi)$  defined in (\ref{least}), and obtain
\bna
\bm\xi_{t+1,i}&=&(\bm\xi_{t+1,i}(1),\cdots,\bm\xi_{t+1,i}(m))^T\nonumber\\
&\triangleq&\arg\min_{\bm\xi}\bar J_{t+1,i}(\bm\xi),\label{kkk}\\
H_{t+1,i}&\triangleq&\{l=1,\cdots,m|\bm\xi_{t+1,i}(l)=0\}.\label{sparse34}
\ena
\end{algorithm}

In the convex objective function (\ref{sparse4}), different
components in $\bm\xi$ are assigned different weights, which is an adaptive LASSO estimator since
 the weights ${\alpha_{t+1,i}}/{\hat{\bm \theta}_{t+1,i}(l)}$ are generated from the local observation sequence
$\{\bm\varphi_{k,j}, y_{k+1,j}, j\in N_{i}\}^t_{k=1}$.
The $\hat{\bm \theta}_{t+1,i}(l)$ appearing in the denominator satisfies that $|\hat{\bm \theta}_{t+1,i}(l)|\geq \sqrt{\frac{\log(\lambda_{\max}(\bm P^{-1}_{t+1,i}))}{\lambda_{\min}(\bm P^{-1}_{t+1,i})}}>0$, which makes (\ref{sparse4}) well defined. Moreover, if $\hat{\bm \theta}_{t+1,i}(l)\rightarrow 0$ for some $l=1,\cdots,m$ and hence $1/\hat{\bm \theta}_{t+1,i}(l)\rightarrow \infty$, then the corresponding minimizer $\bm\xi_{t+1,i}(l)$ should be exactly zero.
This provides an intuitive explanation for the sparse solution of Algorithm \ref{algorithm2} with a finite number of observations.  The set $H_{t+1,i}$ generated from the convex optimization problem (\ref{kkk}) serves as the estimate
for the set $H^*$ defined in  (\ref{sparse5}).
There exist some typical algorithms such as basic pursuit and interior-point  algorithms to solve the convex optimization problem (\ref{kkk}) in the literature
(see e.g., \cite{kim,Gill}).

We introduce the following cooperative non-persistent excitation condition to study the convergence of the sets of zero elements in the unknown sparse parameter vector  with a finite number of observations, which is different from the asymptotic analysis given in the last two subsections.
\begin{assumption} (Cooperative Non-Persistent Excitation  Condition) \label{a4}
The following condition is satisfied,
\bna
\frac{r_t}{\lambda^{n,t}_{\min}}\sqrt{\frac{\log(r_t)}{\lambda^{n,t}_{\min}}}\xrightarrow[{t\rightarrow\infty}] {} 0, {~~~\rm a.s.}\label{cooperative}
\ena
where $r_t$ and $\lambda^{n,t}_{\min}$ are respectively  defined in Theorem \ref{theorem1} and Remark \ref{remark_two}.
\end{assumption}

\begin{remark}
For the single sensor case with $n=1$ and $  D_{\mathcal{G}}=1$,  the condition (\ref {cooperative}) reduces to the
excitation condition given by  \cite{zhao}.
Assumption
\ref{a4} reveals the cooperative effect of multiple sensors in the sense that the condition (\ref{cooperative}) can make it possible for Algorithm \ref{algorithm2} to estimate the unknown parameter $\bm\theta$ and the sets of zero elements 
by the cooperation of
multiple sensors even if any individual sensor cannot due to lack of adequate excitation, which is also shown in the simulation example given in Section \ref{simulaiton}.
\end{remark}

For the set $H_{t,i}$ obtained by (\ref{sparse34}), we get the following finite time convergence result, which shows that the set of zero elements in $\bm\theta$ can
be correctly identified with a finite number of observations.

\begin{theorem}\label{theorem2}
(Set convergence) ~Under Assumptions \ref{a1}-\ref{a4}, if $\log{r_t}=O(\log{r_{t-D_{\mathcal{G}}+1}})$, then there exists a positive
integer $T_0$ (which may depend on the sample $\omega$) such that  for all $i\in\{1,\cdots,n\}$
\ban
\bm\xi_{t+1,i}(d+1)=\cdots=\bm\xi_{t+1,i}(m)=0,  ~~t\geq T_0.
\ean
That is, $H_{t+1,i}=H^*$ for $t\geq T_0$,  where $H^*$ and $H_{t+1,i}$  are defined in (\ref{sparse5}) and (\ref{sparse34}).
\end{theorem}
The detailed
proof of Theorem \ref{theorem2} is given in Subsection \ref{proof:Theorem3}.

\begin{remark}
From Theorem \ref{theorem2} (also Theorem \ref{theorem1} and  Theorem \ref{theorem4} ), we see that
the parameter convergence, regret analysis, and set convergence results in this paper are derived without using the independency
assumption on the regression vectors, which
makes it possible to apply our algorithm to practical feedback systems.
\end{remark}
\section{Proofs of the main results}\label{proofs_of}

In order to prove the main theorems of the paper, we first give two preliminary lemmas.

Denote the estimation error of the classical distributed LS algorithm (\ref{theta1}) as  $\widetilde{\bm \theta}_{t+1,i}\triangleq\bm\theta_{t+1,i}-\bm\theta$, and $\widetilde{\bm \Theta}_t=col\{\widetilde{\bm \theta}_{t,1},...,\widetilde{\bm \theta}_{t,n}\}$.
\begin{lemma}{ \rm\citep{wc2}}\label{lemma1}
Under Assumptions \ref{a1} and \ref{a2},  we have the following results for the classical distributed LS algorithm (\ref{theta1}),
\ban
&1)& \sum^n_{i=1}\|\widetilde{\bm\theta}_{t,i}\|^2=O\left(\frac{\log r_t}{\lambda^{n,t}_{\min}}\right),\\
&2)& \sum^t_{k=0}\lambda_{\max}(\bm d_k\bm \Phi^T_k\bm P_k\bm \Phi_k)=O(\log r_t),\\
&3)& \sum^t_{k=0}\widetilde{\bm \Theta}^T_k\bm \Phi_{k}\bm d_k\bm \Phi^T_{k}\widetilde{\bm \Theta}_k=O(\log r_t),
\ean
where $\bm P_k$ and $\bm \Phi_{k}$ are defined in Theorem \ref{theorem4}, $r_t\triangleq\max\limits_{1\leq i\leq n}\lambda_{\max}\{\bm P^{-1}_{0,i}\}+\sum^n_{i=1}\sum^t_{k=0}\|\bm\varphi_{k,i}\|^2$ and $\bm d_t\triangleq diag\Big\{\frac{1}{1+\bm\varphi^T_{t,1}\bm P_{t,1}\bm\varphi_{t,1}},...,\frac{1}{1+\bm\varphi^T_{t,n}\bm P_{t,n}\bm\varphi_{t,n}}\Big\}.$
\end{lemma}

The following lemma provides an upper bound for the cumulative
summation of the noises.
\begin{lemma}{ \rm\citep{Gan2022}}\label{lemma3}
Under Assumptions \ref{a1} and \ref{a2}, for any $i\in\{1,...,n\}$, we have
\ban
&&\Bigg\|\bm P^{\frac{1}{2}}_{t,i}\left(\sum^n_{j=1}\sum^{t}_{k=0}a^{(t+1-k)}_{ij}\bm\varphi_{k,j}w_{k+1,j}\right)\Bigg\|= O(\sqrt{\log(r_{t})}).
\ean
\end{lemma}

\subsection{ Proof of Theorem \ref{theorem1}}\label{proof:theorem1}
\begin{proof}
By noting that $\bm\beta_{t+1,i}$ is the minimizer of $ J_{t+1,i}(\bm\beta)$, it follows that
\bna
0&\geq& J_{t+1,i}(\bm\beta_{t+1,i})-J_{t+1,i}(\bm\theta)\nonumber\\
&=& J_{t+1,i}(\widetilde{\bm\beta}_{t+1,i}+\bm\theta)-J_{t+1,i}(\bm\theta).\label{sparse9}
\ena
Since $\bm\theta(j)=0$, $j=d+1,\cdots,m$, by (\ref{model}), (\ref{least2}) and (\ref{penalty}), we have
\bna
&&J_{t+1,i}(\widetilde{\bm\beta}_{t+1,i}+\bm\theta)\nonumber\\
&=&\sum^n_{j=1}\sum^{t}_{k=0}a^{(t+1-k)}_{ij}
[w_{k+1,j}-\widetilde{\bm\beta}^T_{t+1,i}\bm\varphi_{k,j}]^2\nonumber\\
&&+\alpha_{t+1,i}\sum^d_{l=1}|{\widetilde{\bm\beta}_{t+1,i}(l)+\bm\theta(l)}|+\alpha_{t+1,i}\sum^{m}_{l=d+1}|{\widetilde{\bm\beta}_{t+1,i}(l)\nonumber}|\\
&=&\sum^n_{j=1}\sum^{t}_{k=0}a^{(t+1-k)}_{ij}
w^2_{k+1,j}\nonumber\\
&&-2\widetilde{\bm\beta}^T_{t+1,i}\sum^n_{j=1}\sum^{t}_{k=0}a^{(t+1-k)}_{ij}
\bm\varphi_{k,j}w_{k+1,j}\nonumber\\
&&+\widetilde{\bm\beta}^T_{t+1,i}\sum^{t}_{k=0}a^{(t+1-k)}_{ij}\bm\varphi_{k,j}
\bm\varphi^T_{k,j}\widetilde{\bm\beta}_{t+1,i}
\nonumber\\
&&+\alpha_{t+1,i}\sum^d_{l=1}|{\widetilde{\bm\beta}_{t+1,i}(l)+\bm\theta(l)}|+\alpha_{t+1,i}\sum^{m}_{l=d+1}|{\widetilde{\bm\beta}_{t+1,i}(l)}|.\nonumber\\
\label{sparse7}
\ena
Similarly, we have
\bna
&&J_{t+1,i}(\bm\theta)\nonumber\\
&=&\sum^n_{j=1}\sum^{t}_{k=0}a^{(t+1-k)}_{ij}
[y_{k+1,j}-{\bm\theta}^T\bm\varphi_{k,j}]^2+\alpha_{t+1,i}\sum^d_{l=1}|\bm\theta(l)|\nonumber\\
&=&\sum^n_{j=1}\sum^{t}_{k=0}a^{(t+1-k)}_{ij}
w^2_{k+1,j}+\alpha_{t+1,i}\sum^d_{l=1}|\bm\theta(l)|.\label{sparse8}
\ena
Hence by (\ref{sparse7}) and (\ref{sparse8}), we have
\bna
&&J_{t+1,i}(\widetilde{\bm\beta}_{t+1,i}+\bm\theta)-J_{t+1,i}(\bm\theta)\nonumber\\
&\geq&\widetilde{\bm\beta}^T_{t+1,i}\sum^{t}_{k=0}a^{(t+1-k)}_{ij}\bm\varphi_{k,j}
\bm\varphi^T_{k,j}\widetilde{\bm\beta}_{t+1,i}\nonumber\\
&&-2\widetilde{\bm\beta}^T_{t+1,i}\sum^n_{j=1}\sum^{t}_{k=0}a^{(t+1-k)}_{ij}
\bm\varphi_{k,j}w_{k+1,j}\nonumber\\
&&+\alpha_{t+1,i}\sum^d_{l=1}(|{\widetilde{\bm\beta}_{t+1,i}(l)+\bm\theta(l)}|-|\bm\theta(l)|)
\nonumber\\
&\triangleq& M^{(1)}_{t+1,i}-2M^{(2)}_{t+1,i}+M^{(3)}_{t+1,i}.\label{sparse10}
\ena
In the following, we estimate $M^{(1)}_{t+1,i}$, $M^{(2)}_{t+1,i}$ and $M^{(3)}_{t+1,i}$ separately.
Denote $\bm V_{t+1,i}=\bm P^{-\frac{1}{2}}_{t+1,i}\widetilde{\bm\beta}_{t+1,i}$.
By Lemma \ref{lemma3}, we have
\ban &&
|M^{(2)}_{t+1,i}|\nonumber\\ &=&\Big|\widetilde{\bm\beta}^T_{t+1,i}\bm P^{-\frac{1}{2}}_{t+1,i}\bm P^{\frac{1}{2}}_{t+1,i}\sum^n_{j=1}\sum^{t}_{k=0}a^{(t+1-k)}_{ij}
\bm\varphi_{k,j}w_{k+1,j}\Big|\nonumber\\
&=&O\big(\sqrt{\log(r_{t})}\big)\|\bm V_{t+1,i}\|.
\ean
Hence, there exists a positive constant $c_1$ such that for large $t$,
\bna
&&M^{(1)}_{t+1,i}-2M^{(2)}_{t+1,i}\nonumber\\
&\geq& \frac{1}{2}\|\bm V_{t+1,i}\|^2-c_1\sqrt{\log(r_{t})}\|\bm V_{t+1,i}\|.
\label{sparse11}
\ena
By $C_r$-inequality, we have
\bna
|M^{(3)}_{t+1,i}|\leq  \alpha_{t+1,i}\sum^d_{l=1}|\widetilde{\bm\beta}_{t+1,i}(l)|
\leq \alpha_{t+1,i}\sqrt{d}\|\widetilde{\bm\beta}_{t+1,i}\|. \label{sparse12}
\ena
Hence by (\ref{sparse9}) and (\ref{sparse10})-(\ref{sparse12}), we have for large $t$
\ban
0\geq \frac{\|\bm V_{t+1,i}\|^2}{2}-c_1\sqrt{\log(r_{t})}\|\bm V_{t+1,i}\|-\sqrt{d}\alpha_{t+1,i}\|\widetilde{\bm\beta}_{t+1,i}\|,
\ean
which implies that
\bna
\|\bm V_{t+1,i}\|\leq \sqrt{c^2_1\log r_t+2\sqrt{d}\alpha_{t+1,i}\|\widetilde{\bm\beta}_{t+1,i}\|}
+\sqrt{c_1\log r_t}.\nonumber\\ \label{sparse13}
\ena
Note that by the definition of  $\bm V_{t+1,i}$, we have
\begin{gather*}
\|\bm V_{t+1,i}\|^2\geq \lambda_{\min}(\bm P^{-1}_{t+1,i})\|\widetilde{\bm\beta}_{t+1,i}\|^2.
\end{gather*}
Combining this with (\ref{sparse13}), we have
\ban
&&\left(\|\widetilde{\bm\beta}_{t+1,i}\|-\frac{2\sqrt{d}\alpha_{t+1,i}}
{\lambda_{\min}(\bm P^{-1}_{t+1,i})}\right)^2\nonumber\\
&\leq &\left(\frac{2\sqrt{d}\alpha_{t+1,i}}
{\lambda_{\min}(\bm P^{-1}_{t+1,i})}\right)^2+\frac{(2c_1^2+2c_1)\log r_t}{\lambda_{\min}(\bm P^{-1}_{t+1,i})}.
\ean
Thus, we have
\bna
\|\widetilde{\bm\beta}_{t+1,i}\|=O\left(\frac{\alpha_{t+1,i}}
{\lambda_{\min}(\bm P^{-1}_{t+1,i})}+\sqrt{\frac{\log r_t}{\lambda_{\min}(\bm P^{-1}_{t+1,i})}}\right),
\ena
which completes the proof of the theorem.
\end{proof}

\subsection{Proof of Theorem \ref{theorem4}}\label{proof:theorem2}

\begin{proof}
By (\ref{least2}), we obtain the subdifferential of (\ref{penalty}),
\ban
\partial J_{t+1,i}(\bm\beta)&=&-2\sum^n_{j=1}\sum^{t}_{k=0}a^{(t+1-k)}_{ij}
\bm\varphi_{k,j}(y_{k+1,j}-\bm\varphi^T_{k,j}{\bm\beta})\\
&&+\alpha_{t+1,i}\partial \|\bm \beta\|_1,
\ean
where $\partial \|\bm \beta\|_1$ is the subdifferential  of $\|\bm \beta\|_1$.
Since $\bm\beta_{t+1,i}$ is the minimizer of $J_{t+1,i}(\bm\beta)$, we have
$\bm 0\in \partial J_{t+1,i}(\bm\beta_{t+1,i})$ with $\bm 0\triangleq (\underbrace{0,0,\cdots,0}_{m})^T$, i.e.,
\bna
\bm 0&\in &-2\sum^n_{j=1}\sum^{t}_{k=0}a^{(t+1-k)}_{ij}
\bm\varphi_{k,j}(y_{k+1,j}-\bm\varphi^T_{k,j}{\bm\beta}_{t+1,i})\nonumber\\
&&+\alpha_{t+1,i}\partial \|\bm \beta_{t+1,i}\|_1.\label{J}
\ena
Let us write (\ref{J}) in a component form, i.e., for all $l\in\{1,\cdots,m\}$,
\bna
0&\in& -\sum^n_{j=1}\sum^{t}_{k=0}a^{(t+1-k)}_{ij}
\bm\varphi_{k,j}(l)y_{k+1,j}\nonumber\\
&&+\sum^n_{j=1}\sum^{t}_{k=0}a^{(t+1-k)}_{ij}
\bm\varphi_{k,j}(l)\Big(\sum_{s\neq l}\bm\varphi_{k,j}(s){\bm\beta_{t+1,i}}(s)\Big)\nonumber\\
&&+\sum^n_{j=1}\sum^{t}_{k=0}a^{(t+1-k)}_{ij}
\bm\varphi^2_{k,j}(l){\bm\beta_{t+1,i}}(l)+\frac{\alpha_{t+1,i}}{2}\partial |\bm \beta_{t+1,i}(l)|\nonumber\\
&\triangleq &\bm D_{t+1,i}(l)+\sum^n_{j=1}\sum^{t}_{k=0}a^{(t+1-k)}_{ij}
\bm\varphi^2_{k,j}(l){\bm\beta_{t+1,i}}(l)\nonumber\\
&&+\frac{\alpha_{t+1,i}}{2}\partial |\bm \beta_{t+1,i}(l)|.\label{classify}
\ena

Note that
\begin{alignat}{2}
\partial |\bm \beta_{t+1,i}(l)|=\left\{
\begin{aligned}
&1, ~~~~~~~~~~~~~~~~~~{\rm if} ~~~\bm\beta_{t+1,i}(l)>0\\
&-1, ~~~~~~~~~~~~~~{\rm if} ~~~\bm \beta_{t+1,i}(l)<0\\
&\in [-1,1],~~~~~~{\rm if} ~~~\bm \beta_{t+1,i}(l)=0\nonumber
\end{aligned}\ .
\right.
\end{alignat}
Set $\sum^n_{j=1}\sum^t_{k=0}a^{(t+1-k)}_{ij}\bm\varphi_{k,j}\bm\varphi^T_{k,j}
\triangleq\bm \Psi_{t+1,i}$. Combining the above equation with (\ref{classify}) yields for large $t$
\begin{alignat}{2} &
\bm \beta_{t+1,i}(l)\nonumber\\ =&\left\{
\begin{aligned}
&\frac{-\bm D_{t+1,i}(l)+\frac{\alpha_{t+1,i}}{2} }{\bm \Psi_{t+1,i}(l,l)},~{\rm if} ~\bm D_{t+1,i}(l)>\frac{\alpha_{t+1,i}}{2}\\
&\frac{-\bm D_{t+1,i}(l)-\frac{\alpha_{t+1,i}}{2}}{\bm \Psi_{t+1,i}(l,l)},~{\rm if} ~\bm D_{t+1,i}(l)<-\frac{\alpha_{t+1,i}}{2}\\
&~~~ 0,
\hskip 2.6cm
{\rm if}~|\bm D_{t+1,i}(l)|\leq\frac{\alpha_{t+1,i}}{2}\nonumber
\end{aligned}\ \ ,
\right.
\end{alignat}
with $\bm \Psi_{t+1,i}(l,l)$ being the   $l$-th  diagonal element of the matrix  $\bm \Psi_{t+1,i}$.
This  implies that
\bna
\bm \Psi_{t+1,i}(l,l)\bm \beta_{t+1,i}(l)=
-\bm D_{t+1,i}(l)+\bm \gamma_{t+1,i}(l),\label{sparse1}
\ena
where
\begin{alignat}{2}
\bm \gamma_{t+1,i}(l)=\left\{
\begin{aligned}
&\frac{\alpha_{t+1,i}}{2}, \hskip 1.4cm{\rm if}~\bm D_{t+1,i}(l)>\frac{\alpha_{t+1,i}}{2}\\
&-\frac{\alpha_{t+1,i}}{2}, \hskip 1.0cm{\rm if} ~\bm D_{t+1,i}(l)<-\frac{\alpha_{t+1,i}}{2}\\
&\bm D_{t+1,i}(l),\hskip 1.0cm{\rm if}~|\bm D_{t+1,i}(l)|\leq\frac{\alpha_{t+1,i}}{2}\nonumber
\end{aligned}.
\right.
\end{alignat}
Then by (\ref{sparse1}) and the definition of $\bm D_{t+1,i}(l)$, we have for all $l\in\{1,\cdots,m\}$
\ban
&&\sum^n_{j=1}\sum^{t}_{k=0}a^{(t+1-k)}_{ij}
\bm\varphi_{k,j}(l)\bm\varphi^T_{k,j}{\bm\beta_{t+1,i}}\\
&=&\sum^n_{j=1}\sum^{t}_{k=0}a^{(t+1-k)}_{ij}
\bm\varphi_{k,j}(l)y_{k+1,j}+\bm \gamma_{t+1,i}(l).
\ean
We rewrite the above equation into the matrix form, and obtain the following equation by  (\ref{theta}) for large $t$
\bna
\bm\beta_{t+1,i}&=&\bm P_{t+1,i}\left(\sum^n_{j=1}\sum^{t}_{k=0}a^{(t+1-k)}_{ij}
\bm\varphi_{k,j}y_{k+1,j}+\bm \gamma_{t+1,i}\right)\nonumber\\
&=&\bm\theta_{t+1,i}+\bm P_{t+1,i}\bm \gamma_{t+1,i}\label{sparse2},
\ena
where $\bm \gamma_{t+1,i}=(\bm \gamma_{t+1,i}(1),\cdots,\bm \gamma_{t+1,i}(m))^T$ and $\bm\theta_{t+1,i}$ is defined in (\ref{theta}).
 Note that for all $i\in\{1,\cdots,n\}$ and
$l\in\{1,\cdots,m\}$,
$\|\bm \gamma_{t+1,i}(l)\|\leq\frac{\alpha_{t+1,i}}{2}$, hence by  Lemma \ref{lemma1}, we obtain
\bna
&&\sum^t_{k=0}\bm\gamma^T_k\bm P_k\bm \Phi_{k}\bm d_k\bm\Phi^T_{k}\bm P_k\bm\gamma_k\nonumber\\
&\leq&\sum^t_{k=0}\lambda_{\max}(\bm d_k\bm \Phi^T_{k}\bm P_k\bm \Phi_{k})\bm\gamma^T_k\bm P_k\bm\gamma_k\nonumber\\
&\leq&\sum^t_{k=0}\left[\lambda_{\max}(\bm d_k\bm \Phi^T_{k}\bm P_k\bm \Phi_{k})\left(\sum^n_{i=1}\lambda_{\max}(\bm P_{k,i})\|\bm\gamma_{k,i}\|^2\right)\right]\nonumber\\
&=&O\left(\sum^t_{k=0}\left[\lambda_{\max}(\bm d_k\bm \Phi^T_{k}\bm P_k\bm \Phi_{k})
\left(\sum^n_{i=1}\frac{\alpha^2_{k,i}}{\lambda_{\min}(\bm P^{-1}_{k,i})}\right)\right]\right)\nonumber\\
&=&O(\log r_t),\label{gamma}
\ena
where $\bm \gamma_{t+1}=col\{\bm \gamma_{t+1,1},\cdots,\bm \gamma_{t+1,n}\}$.
By the definition of $\bm d_t$ in Lemma \ref{lemma1},
 we have $\bm I_n=\bm d_k+\bm d_k\bm \Phi^T_{k}\bm P_k\bm \Phi_{k}$. By (\ref{sparse2}), we have $\widetilde{\bm \beta}_{t+1}=\widetilde{\bm\Theta}_{t+1}+\bm P_{t+1}\bm \gamma_{t+1}$,  where $\widetilde{\bm \beta}_t=col\{\widetilde{\bm \beta}_{t,1},...,\widetilde{\bm \beta}_{t,n}\}$. Hence
by (\ref{gamma}), Lemma \ref{lemma1} and the condition $\bm\Phi^T_{t}\bm P_t\bm\Phi_{t}=O(1)$, we have
\ban
&&R_t=\sum^t_{k=0}\widetilde{\bm \beta}^T_k\bm \Phi_{k}\bm \Phi^T_{k}\widetilde{\bm \beta}_k\nonumber\\
&=&\sum^t_{k=0}\widetilde{\bm \beta}^T_k\bm \Phi_{k}\bm d_k\bm \Phi^T_{k}\widetilde{\bm \beta}_k
+\sum^t_{k=0}\widetilde{\bm \beta}^T_k\bm \Phi_{k}(\bm d_k\bm \Phi^T_{k}\bm P_k\bm \Phi_{k})\bm \Phi^T_{k}\widetilde{\bm \beta}_k\nonumber\\
&=&O\Big(\sum^t_{k=0}\widetilde{\bm \beta}^T_k\bm \Phi_{k}\bm d_k\bm\Phi^T_{k}\widetilde{\bm \beta}_k\Big)\\
&=&O\Big(\sum^t_{k=0}\widetilde{\bm \Theta}^T_k\bm \Phi_{k}\bm d_k\bm \Phi^T_{k}\widetilde{\bm \Theta}_k+\sum^t_{k=0}\bm\gamma^T_k\bm P_k\bm \Phi_{k}\bm d_k\bm\Phi^T_{k}\bm P_k\bm\gamma_k\Big)\\
&=&O(\log r_t).
\ean
This completes the proof of the theorem.
\end{proof}

\subsection{Proof of Theorem \ref{theorem2} } \label{proof:Theorem3}
\begin{proof}
Denote the estimation error between $\bm\xi_{t+1,i}$ obtained by Algorithm \ref{algorithm2} and $\bm\theta$  as
\begin{gather}
\widetilde{\bm\xi}_{t+1,i}=\bm\xi_{t+1,i}-\bm\theta.\label{sparse36}
\end{gather}
By Assumption \ref{a4} and Lemma \ref{lemma1}, we see that the limits of
$\bm \theta_{t+1,i}(l)$ and $\hat{\bm \theta}_{t+1,i}(l)$, $l=1,\cdots,d$ are nonzero. Similar to the proof
of Theorem \ref{theorem1}, we also have the following result,
\bna
\|\widetilde{\bm\xi}_{t+1,i}\|=O\left(\frac{\alpha_{t+1,i}}
{\lambda_{\min}(\bm P^{-1}_{t+1,i})}+\sqrt{\frac{\log r_t}{\lambda_{\min}(\bm P^{-1}_{t+1,i})}}\right).\label{sparse35}
\ena
By the definition of $\widetilde{\bm\xi}_{t+1,i}$ in (\ref{sparse36}), it suffices to prove  that
there exists a positive
integer $T_0$ such that for all $i\in\{1,\cdots,n\}$
\ban
\widetilde{\bm\xi}_{t+1,i}(d+1)=\cdots=\widetilde{\bm\xi}_{t+1,i}(m)=0, ~~~~t\geq T_0.
\ean
Otherwise, if for some $s_l\in\{d+1,\cdots,m\}$, some sensor $i_0$, and some subsequence $\{t_p\}_{p\geq 1}$
such that $\widetilde{\bm\xi}_{t_p+1,i_0}(s_l)\neq 0$, $p\geq 1$. Thus for $p\geq 1$, we have  $\|\widetilde{\bm\xi}_{t_p+1,i_0}\|>0$.

Denote
\begin{align}
\widetilde{\bm\xi}_{t_p+1,i_0}&=\left(
\begin{matrix}
\widetilde{\bm\xi}^{(1)}_{t_p+1,i_0}\\
\widetilde{\bm\xi}^{(2)}_{t_p+1,i_0}\\
\end{matrix}
\right)~  {\rm and}~
\bar{\bm\xi}_{t_p+1,i_0}=\left(
\begin{matrix}
\widetilde{\bm\xi}^{(1)}_{t_p+1,i_0}\\
\bm 0\\
\end{matrix}
\right),\label{sparse17}
\end{align}
where $\widetilde{\bm\xi}^{(1)}_{t_p+1,i_0}
\in \mathbb{R}^d$  and $\widetilde{\bm\xi}^{(2)}_{t_p+1,i_0}
\in \mathbb{R}^{m-d}$.  By noting that $\bm\xi_{t_p+1,i_0}$ is the minimizer of $ \bar J_{t_p+1,i_0}(\bm\xi)$ defined by (\ref{sparse4}) ,
it follows that
\bna
0&\geq &\bar J_{t_p+1,i_0}(\bm\xi_{t_p+1,i_0})- \bar J_{t_p+1,i_0}(\bm\theta+\bar{\bm\xi}_{t_p+1,i_0})\nonumber\\
&=&\bar J_{t_p+1,i_0}(\bm\theta+\widetilde{\bm\xi}_{t_p+1,i_0})- \bar J_{t_p+1,i_0}(\bm\theta+\bar{\bm\xi}_{t_p+1,i_0}).
\label{sparse29}
\ena
Denote
\begin{align}
&\bm \Psi_{t+1,i}=\sum^n_{j=1}\sum^t_{k=0}a^{(t+1-k)}_{ij}\bm\varphi_{k,j}\bm\varphi^T_{k,j}
\triangleq\left(
\begin{matrix}
\bm \Psi^{(11)}_{t+1,i}&\bm \Psi^{(12)}_{t+1,i}\\
\bm \Psi^{(21)}_{t+1,i}&\bm \Psi^{(22)}_{t+1,i}\\
\end{matrix}
\right),\nonumber\\
&~~{\rm and }~~
\bm \varphi_{k,j}\triangleq\left(
\begin{matrix}
\bm \varphi^{(1)}_{k,j}\\
\bm \varphi^{(2)}_{k,j}\\
\end{matrix}
\right).\label{sparse20}
\end{align}
Similar to (\ref{sparse7}), we have for $\widetilde{\bm\xi}_{t_p+1,i_0}$
\bna
&&\bar J_{t_p+1,i_0}(\bm\theta+\widetilde{\bm\xi}_{t_p+1,i_0})-\sum^n_{j=1}\sum^{t_p}_{k=0}a^{(t_p+1-k)}_{i_0j}w^2_{k+1,j}\nonumber\\
&=&-2\widetilde{\bm\xi}^{(1)T}_{t_p+1,i_0}\sum^n_{j=1}\sum^{t_p}_{k=0}a^{(t_p+1-k)}_{i_0j}
\bm\varphi^{(1)}_{k,j}w_{k+1,j}\nonumber\\
&&-2\widetilde{\bm\xi}^{(2)T}_{t_p+1,i_0}\sum^n_{j=1}\sum^{t_p}_{k=0}a^{(t_p+1-k)}_{i_0j}
\bm\varphi^{(2)}_{k,j}w_{k+1,j}\nonumber\\
&&+\widetilde{\bm\xi}^{(1)T}_{t_p+1,i_0}\bm \Psi^{(11)}_{t_p+1,i_0}\widetilde{\bm\xi}^{(1)}_{t_p+1,i_0}+\widetilde{\bm\xi}^{(2)T}_{t_p+1,i_0}\bm \Psi^{(21)}_{t_p+1,i_0}\widetilde{\bm\xi}^{(1)}_{t_p+1,i_0}\nonumber\\
&&
+\widetilde{\bm\xi}^{(1)T}_{t_p+1,i_0}\bm \Psi^{(12)}_{t_p+1,i_0}\widetilde{\bm\xi}^{(2)}_{t_p+1,i_0}
+\widetilde{\bm\xi}^{(2)T}_{t_p+1,i_0}\bm \Psi^{(22)}_{t_p+1,i_0}\widetilde{\bm\xi}^{(2)}_{t_p+1,i_0}\nonumber\\
&&+\alpha_{t_p+1,i_0}\sum^d_{l=1}\frac{1}{\hat{\bm \theta}_{t_p+1,i_0}(l)}|{\widetilde{\bm\xi}_{t_p+1,i_0}(l)+\bm\theta(l)}|\nonumber\\
&&+\alpha_{t_p+1,i_0}\sum^{m}_{l=d+1}\frac{1}{|\hat{\bm \theta}_{t_p+1,i_0}(l)|}|{\widetilde{\bm\xi}_{t_p+1,i_0}(l)}|.\label{sparse18}
\ena
For $\bar{\bm\xi}_{t_p+1,i_0}$ defined in (\ref{sparse17}), we have
\bna
&&\bar J_{t_p+1,i_0}(\bm\theta+\bar{\bm\xi}_{t_p+1,i_0})-\sum^n_{j=1}\sum^{t_p}_{k=0}a^{(t_p+1-k)}_{i_0j}w^2_{k+1,j}\nonumber\\
&=&-2\widetilde{\bm\xi}^{(1)T}_{t_p+1,i_0}\sum^n_{j=1}\sum^{t_p}_{k=0}a^{(t_p+1-k)}_{i_0j}
\bm\varphi^{(1)}_{k,j}w_{k+1,j}\nonumber\\
&&+\widetilde{\bm\xi}^{(1)T}_{t_p+1,i_0}\bm \Psi^{(11)}_{t_p+1,i_0}\widetilde{\bm\xi}^{(1)}_{t_p+1,i_0}\nonumber\\
&&+\alpha_{t_p+1,i_0}\sum^d_{l=1}\frac{1}{|\hat{\bm \theta}_{t_p+1,i_0}(l)|}|{\bar{\bm\xi}_{t_p+1,i_0}(l)+\bm\theta(l)}|.\label{sparse19}
\ena
By (\ref{sparse18}) and (\ref{sparse19}), we have
\bna
&&\bar J_{t_p+1,i_0}(\bm\theta+\widetilde{\bm\xi}_{t_p+1,i_0})-\bar J_{t_p+1,i_0}(\bm\theta+\bar{\bm\xi}_{t_p+1,i_0})\nonumber\\
&=&-2\widetilde{\bm\xi}^{(2)T}_{t_p+1,i_0}\sum^n_{j=1}\sum^{t_p}_{k=0}a^{(t_p+1-k)}_{i_0j}
\bm\varphi^{(2)}_{k,j}w_{k+1,j}\nonumber\\
&&+\widetilde{\bm\xi}^{(2)T}_{t_p+1,i_0}\bm \Psi^{(22)}_{t_p+1,i_0}\widetilde{\bm\xi}^{(2)}_{t_p+1,i_0}+\widetilde{\bm\xi}^{(1)T}_{t_p+1,i_0}\bm \Psi^{(12)}_{t_p+1,i_0}\widetilde{\bm\xi}^{(2)}_{t_p+1,i_0}\nonumber\\
&&+\widetilde{\bm\xi}^{(2)T}_{t_p+1,i_0}\bm \Psi^{(21)}_{t_p+1,i_0}\widetilde{\bm\xi}^{(1)}_{t_p+1,i_0}\nonumber\\
&&+\alpha_{t_p+1,i_0}\sum^{m}_{l=d+1}\frac{1}{|\hat{\bm \theta}_{t_p+1,i_0}(l)|}|{\widetilde{\bm\xi}_{t_p+1,i_0}(l)}|\nonumber\\
&\triangleq&  -2I^{(1)}_{t_p+1,i_0}+I^{(2)}_{t_p+1,i_0}+I^{(3)}_{t_p+1,i_0}+I^{(4)}_{t_p+1,i_0}+I^{(5)}_{t_p+1,i_0}.\nonumber\\ \label{sparse26}
\ena
In the following, we estimate $I^{(1)}_{t_p+1,i_0}$, $I^{(2)}_{t_p+1,i_0}$, $I^{(3)}_{t_p+1,i_0}$, $I^{(4)}_{t_p+1,i_0}$, $I^{(5)}_{t_p+1,i_0}$ separately.
By (\ref{sparse16}) and (\ref{sparse20}), we have
\begin{align*}
\bm P^{-1}_{t+1,i}&=\bm \Psi_{t+1,i}
+\sum^n_{j=1}a^{(t+1)}_{ij}\bm P^{-1}_{0,j}\triangleq\left(
\begin{matrix}
\bm Q^{(11)}_{t+1,i}&\bm Q^{(12)}_{t+1,i}\\
\bm Q^{(21)}_{t+1,i}&\bm Q^{(22)}_{t+1,i}\\
\end{matrix}
\right).
\end{align*}
By (\ref{sparse20}) and Lemma \ref{lemma3}, we have
\ban
|I^{(1)}_{t_p+1,i_0}|&=&\Bigg|\widetilde{\bm\xi}^{(2)T}_{t_p+1,i_0}(\bm Q^{(22)}_{t_p+1,i_0})^{\frac{1}{2}}
(\bm Q^{(22)}_{t_p+1,i_0})^{-\frac{1}{2}}\\
&&\sum^n_{j=1}\sum^{t_p}_{k=0}a^{(t_p+1-k)}_{i_0j}
\bm\varphi^{(2)}_{k,j}w_{k+1,j}\Bigg|\\
&=&\|(\bm Q^{(22)}_{t_p+1,i_0})\|^{\frac{1}{2}}\|\widetilde{\bm\xi}^{(2)}_{t_p+1,i_0}\|O\left(\sqrt{\log r^{(2)}_{t_p}}\right),
\ean
where $r^{(2)}_{t}\triangleq \max\limits_{1\leq i\leq n}\lambda_{\max}\{\bm Q^{(22)}_{0,i}\}+\sum^n_{i=1}\sum^t_{k=0}\|\bm\varphi^{(2)}_{k,i}\|^2$.

Note that  $\lambda_{\max}(\bm Q^{(22)}_{t_p+1,i_0})\leq
 \lambda_{\max}(\bm P^{-1}_{t_p+1,i_0})$ and $\lambda_{\min}(\bm Q^{(22)}_{t_p+1,i_0})\geq
 \lambda_{\min}(\bm P^{-1}_{t_p+1,i_0})$. Hence, we have $r^{(2)}_{t_p}\leq r_{t_p}$. We obtain that for large $p$ and some positive constant $c_2$
\bna
&&-2I^{(1)}_{t_p+1,i_0}+I^{(2)}_{t_p+1,i_0}\nonumber\\
&\geq&\lambda_{\min}(\bm \Psi^{(22)}_{t_p+1,i_0})\|\widetilde{\bm\xi}^{(2)}_{t_p+1,i_0}\|^2\nonumber\\
&&-c_2\|(\bm Q^{(22)}_{t_p+1,i_0})\|^{\frac{1}{2}}\|\widetilde{\bm\xi}^{(2)}_{t_p+1,i_0}\|\sqrt{\log r^{(2)}_{t_p}}\nonumber\\
&\geq&\frac{1}{2}\lambda_{\min}(\bm Q^{(22)}_{t_p+1,i_0})\|\widetilde{\bm\xi}^{(2)}_{t_p+1,i_0}\|^2\nonumber\\
&&-c_2\|(\bm Q^{(22)}_{t_p+1,i_0})\|^{\frac{1}{2}}\|\widetilde{\bm\xi}^{(2)}_{t_p+1,i_0}\|\sqrt{\log r^{(2)}_{t_p}}\nonumber\\
&\geq&\frac{1}{2}\lambda_{\min}(\bm P^{-1}_{t_p+1,i_0})\|\widetilde{\bm\xi}^{(2)}_{t_p+1,i_0}\|^2\nonumber\\
&&-c_2\sqrt{\lambda_{\max}(\bm P^{-1}_{t_p+1,i_0})}\|\widetilde{\bm\xi}^{(2)}_{t_p+1,i_0}\|\sqrt{\log r_{t_p}}.\label{sparse22}
\ena

By (\ref{sparse35}) and Lemma \ref{lemma1}, and based on the equivalence of norms in a finite dimensional space, we have
\bna
|I^{(3)}_{t_p+1,i_0}|&=&|\widetilde{\bm\xi}^{(1)T}_{t_p+1,i_0}\bm \Psi^{(12)}_{t_p+1,i_0}\widetilde{\bm\xi}^{(2)}_{t_p+1,i_0}|\nonumber\\
&\leq &\|\widetilde{\bm\xi}^{(1)}_{t_p+1,i_0}\|\|\bm \Psi^{(12)}_{t_p+1,i_0}\|\|\widetilde{\bm\xi}^{(2)}_{t_p+1,i_0}\|\nonumber\\
&\leq& c_3\|\widetilde{\bm\xi}^{(1)}_{t_p+1,i_0}\|\|\widetilde{\bm\xi}^{(2)}_{t_p+1,i_0}\|
\|\bm \Psi^{(12)}_{t_p+1,i_0}\|_F\nonumber\\
&\leq& c_3\|\widetilde{\bm\xi}^{(1)}_{t_p+1,i_0}\|\|\widetilde{\bm\xi}^{(2)}_{t_p+1,i_0}\|
\|\bm \Psi_{t_p+1,i_0}\|_F\nonumber\\
&\leq& c_4\|\widetilde{\bm\xi}^{(1)}_{t_p+1,i_0}\|\|\widetilde{\bm\xi}^{(2)}_{t_p+1,i_0}\|
\|\bm \Psi_{t_p+1,i_0}\|\nonumber\\
&\leq& c_4\|\widetilde{\bm\xi}^{(1)}_{t_p+1,i_0}\|\|\widetilde{\bm\xi}^{(2)}_{t_p+1,i_0}\|
\lambda_{\max}(\bm P^{-1}_{t_p+1,i_0})\nonumber\\
&=&O\Bigg(\lambda_{\max}(\bm P^{-1}_{t_p+1,i_0})\Bigg[\frac{\alpha_{t_p+1,i_0}}
{\lambda_{\min}(\bm P^{-1}_{t_p+1,i_0})}\nonumber\\
&&+\sqrt{\frac{\log(r_{t_p})}{\lambda_{\min}(\bm P^{-1}_{t_p+1,i_0})}}\Bigg]\|\widetilde{\bm\xi}^{(2)}_{t_p+1,i_0}\|\Bigg),
\label{sparse23}
\ena
where $c_3$ and $c_4$ are two positive constants.

Similarly, we have
\bna
|I^{(4)}_{t_p+1,i_0}|&\leq&O\Bigg(\lambda_{\max}(\bm P^{-1}_{t_p+1,i_0})\Bigg[\frac{\alpha_{t_p+1,i_0}}
{\lambda_{\min}(\bm P^{-1}_{t_p+1,i_0})}\nonumber\\
&&+\sqrt{\frac{\log(r_{t_p})}{\lambda_{\min}(\bm P^{-1}_{t_p+1,i_0})}}\Bigg]\|\widetilde{\bm\xi}^{(2)}_{t_p+1,i_0}\|\Bigg).
\label{sparse24}
\ena
Then by the definition of $\hat{\bm \theta}_{t_p+1,i_0}(l)$ in (\ref{sparse6}), and the condition $\log{r_t}=O(\log{r_{t-D_{\mathcal{G}}+1}})$,
we have for $l=d+1,\cdots,m$,
\ban
L_{t_p+1,i_0}
\leq |\hat{\bm \theta}_{t_p+1,i_0}(l)|\leq c_5 L_{t_p+1,i_0},
\ean
where $c_5>0$ is a positive constant, and $$L_{t_p+1,i_0}=\sqrt{\frac{\log(\lambda_{\max}(\bm P^{-1}_{t_p+1,i_0}))}{\lambda_{\min}(\bm P^{-1}_{t_p+1,i_0})}}.$$

Hence we have
\bna
I^{(5)}_{t_p+1,i_0}&\geq& \alpha_{t_p+1,i_0}\frac{1}{c_5L_{t_p+1,i_0}}\sum^{m}_{l=d+1}|{\widetilde{\bm\xi}_{t_p+1,i_0}(l)}|\nonumber\\
&\geq& \alpha_{t_p+1,i_0}\frac{1}{c_5L_{t_p+1,i_0}}\|{\widetilde{\bm\xi}^{(2)}_{t_p+1,i_0}}\|.
\label{sparse25}
\ena
Thus, by (\ref{sparse26})-(\ref{sparse25}), for some $c_{6}>0$,  we obtain
\bna
&&\bar J_{t_p+1,i_0}(\bm\theta+\widetilde{\bm\xi}_{t_p+1,i_0})-\bar J_{t_p+1,i_0}
(\bm\theta+\bar{\bm\xi}_{t_p+1,i_0})\nonumber\\
&\geq&\lambda_{\min}(\bm P^{-1}_{t_p+1,i_0})\|\widetilde{\bm\xi}^{(2)}_{t_p+1,i_0}\|\cdot\nonumber\\
&&
\left(\frac{\|\widetilde{\bm\xi}^{(2)}_{t_p+1,i_0}\|}{2}-c_2\sqrt{\frac{\lambda_{\max}(\bm P^{-1}_{t_p+1,i_0})}{\lambda_{\min}(\bm P^{-1}_{t_p+1,i_0})}}\sqrt{\frac{\log r_{t_p}}{\lambda_{\min}(\bm P^{-1}_{t_p+1,i_0})}}\right.\nonumber\\
&&-\frac{c_{6}\lambda_{\max}(\bm P^{-1}_{t_p+1,i_0})}{\lambda_{\min}(\bm P^{-1}_{t_p+1,i_0})}\Bigg[\frac{\alpha_{t_p+1,i_0}}
{\lambda_{\min}(\bm P^{-1}_{t_p+1,i_0})}+\nonumber\\
&&\sqrt{\frac{\log(r_{t_p})}{\lambda_{\min}(\bm P^{-1}_{t_p+1,i_0})}}\Bigg]\left.+\frac{\alpha_{t_p+1,i_0}}{c_5\lambda_{\min}(\bm P^{-1}_{t_p+1,i_0})L_{t_p+1,i_0}}\right).\label{sparse30}
\ena
By (\ref{sparse28}), (\ref{sparse27}) and Assumption \ref{a4}, we have
\bna
&&\frac{\lambda_{\max}(\bm P^{-1}_{t_p+1,i_0})}{\lambda_{\min}(\bm P^{-1}_{t_p+1,i_0})}\sqrt{\frac{\log r_{t_p}}{\lambda_{\min}(\bm P^{-1}_{t_p+1,i_0})}}\nonumber\\
&\leq& \frac{\lambda_{\max}(\bm P^{-1}_{t_p+1,i_0})}{\lambda_{\min}(\bm P^{-1}_{t_p+1,i_0})}\sqrt{\frac{\log r_{t_p}}{\lambda^{n,t_p}_{\min}}}\nonumber\\
&=&
o\left(\frac{\alpha_{t_p+1,i_0}}{\lambda_{\min}(\bm P^{-1}_{t_p+1,i_0})L_{t_p+1,i_0}}\right).\label{sparse31}
\ena
By (\ref{sparse28}) and Assumption \ref{a4}, we have
\bna
&&\frac{\lambda_{\max}(\bm P^{-1}_{t_p+1,i_0})\alpha_{t_p+1,i_0}}{\lambda^2_{\min}(\bm P^{-1}_{t_p+1,i_0})}\Bigg/\frac{\alpha_{t_p+1,i_0}}{\lambda_{\min}(\bm P^{-1}_{t_p+1,i_0})L_{t_p+1,i_0}}\nonumber\\
&=&L_{t_p+1,i_0}\frac{\lambda_{\max}(\bm P^{-1}_{t_p+1,i_0})}{\lambda_{\min}(\bm P^{-1}_{t_p+1,i_0})}\nonumber\\
&=&O\left(\frac{r_{t_p}}{\lambda^{n,{t_p}}_{\min}}\sqrt{\frac{\log(r_{t_p})}{\lambda^{n,{t_p}}_{\min}}}\right)
=o(1).\label{sparse32}
\ena
From  (\ref{sparse30})-(\ref{sparse32}), we have
\bna
&& \bar J_{t_p+1,i_0}(\bm\theta+\widetilde{\bm\xi}_{t_p+1,i_0})-\bar J_{t_p+1,i_0}
(\bm\theta+\bar{\bm\xi}_{t_p+1,i_0})\nonumber\\
&\geq&\lambda_{\min}(\bm P^{-1}_{t_p+1,i_0})\|\widetilde{\bm\xi}^{(2)}_{t_p+1,i_0}\|\cdot\nonumber\\
&&\left(\frac{\|\widetilde{\bm\xi}^{(2)}_{t_p+1,i_0}\|}{2}+\frac{[\frac{1}{c_5}+o(1)]\alpha_{t_p+1,i_0}}{\lambda_{\min}(\bm P^{-1}_{t_p+1,i_0})L_{t_p+1,i_0}}\right).\label{sparse33}
\ena
 Note that $\widetilde{\bm\xi}_{t_p+1,i_0}(s_l)\neq 0$ for some $s_l\in\{d+1,\cdots,m\}$. Hence
$\|\widetilde{\bm\xi}^{(2)}_{t_p+1,i_0}\|>0$.  Then by (\ref{sparse33}), we have
$ J_{t_p+1,i_0}(\bm\theta+\widetilde{\bm\xi}_{t_p+1,i_0})-\bar J_{t_p+1,i_0}
(\bm\theta+\bar{\bm\xi}_{t_p+1,i_0})>0$, which contradicts (\ref{sparse29}). This implies that $\|\widetilde{\bm\xi}^{(2)}_{t+1,i}\|=0$ for all large $t$ and
 all $i\in\{1,\cdots,n\}$. We complete the proof of the theorem.
\end{proof}

\section{A simulation example}\label{simulaiton}

In this section, we provide an example to
illustrate the performance of the distributed sparse identification algorithm (i.e., Algorithm \ref{algorithm2}) proposed in this paper.
\begin{example}
Consider a network composed of  $n=6$ sensors whose dynamics obey the model (\ref{model}) with the dimension $m=5$.
The noise sequence
$\{w_{t,i}, t\geq1, i=1,\cdots,n\}$  in (\ref{model}) is independent and identically distributed with $w_{t,i}
\sim  \mathcal{N}(0, 0.1)$ (Gaussian distribution with zero mean and variance $0.1$).
Let the regression vectors $\bm \varphi_{t,i}\in\mathbb{R}^{m}$ $ ~(i = 1,\cdots,n, ~t\geq1)$ be generated by the following state space model,
\begin{alignat}{2}
\begin{aligned}
&\bm x_{t,i}&=& \bm A_i \bm x_{t-1,i}+\bm B_i \varepsilon_{t,i},\\
&\bm\varphi_{t,i}&=& \bm C_i \bm x_{t,i},
\end{aligned},
\label{cdls2}
\end{alignat}
where $\bm x_{t,i}\in\mathbb{R}^{m}$ is the state of the above system with $\bm x_{0,i}=[\underbrace{1,\cdots,1}_{m}]^T$, the matrices $\bm A_i$, $\bm B_i$ and $\bm C_i$ ($i=1,2,\cdots, n$) are chosen according to the following way such that the regression vector $\bm \varphi_{t,i}$ is lack  of adequate excitation for any individual sensor,
\begin{eqnarray*}
\bm A_i&=&diag\{\underbrace{1.1,\cdots,1.1}_{m}\}, \\
\bm B_i&=&\bm {e_j}\in\mathbb{R}^{m},\\
\bm C_i&=&
col\{0,\cdots,0,\underset{j^{th}}{\bm {e}_j},0,\cdots,0\}\in\mathbb{R}^{m\times  m},\\
\end{eqnarray*}
where $j=\mod(i,m)$ and $\bm e_j ~(j=1,\cdots,m)$ is the $j$th column of the  identity matrix $\bm I_m$ ~$(m=5)$.
Let the noise sequence $\{\varepsilon_{t,i}, t \geq  1, i =1,\cdots, n\}$ in (\ref{cdls2}) be independent and identically distributed with $\varepsilon_{t,i}
\sim  \mathcal{N}(0, 0.2)$.
All sensors will estimate an unknown  parameter $$\bm\theta=[\bm\theta(1),\bm\theta(2),\bm\theta(3),\bm\theta(4),\bm\theta(5)]^T=[0.8,1.6,0,0,0]^T.$$ The initial estimate is taken  as
$\bm\xi_{0,i}=[{1,1,1,1,1}]^T$ for $i=1,2,\cdots, 6$.
We use the Metropolis rule  \citep{new05} to construct
the weights of the network, i.e.,
\begin{alignat}{2} \label{weight}
a_{li}=
\left\{
\begin{aligned}
&1-\sum_{j\neq i}a_{ij}~~~~~~~~~~~{\rm if}~~l=i\\
&1/(\max\{n_i,n_l\})~~{\rm if}~~l\in N_i\setminus\{i\}
\end{aligned},
\right.
\end{alignat}
 where $n_i$ is the degree of the node $i$.
\end{example}

It  can be  verified that for each sensor $i$ $(i=1\cdots,6)$, the
regression signals $\bm \varphi_{t,i}$ ( generated by (\ref{cdls2})) have no adequate excitation to estimate the unknown parameter, but they can cooperate to satisfy
Assumption \ref{a4}. We repeat the simulation for $s = 100$ times
with the same initial states.

\begin{figure}[htb]
\centering
\includegraphics[width=3in]{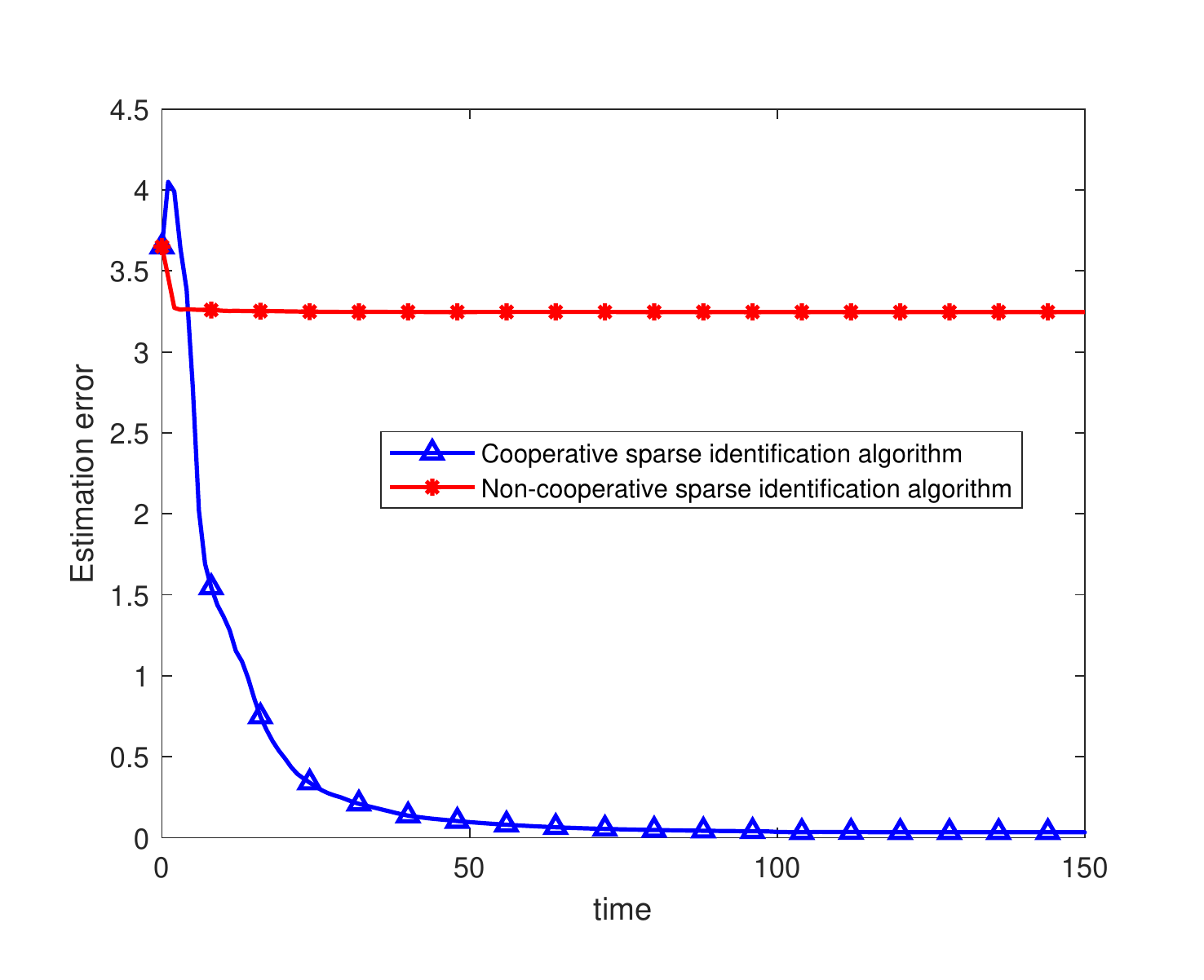}
\caption{The estimation errors of the distributed sparse identification algorithm and non-cooperative sparse identification algorithm }\label{compare}
\end{figure}

\begin{figure*}[htb]
\centering
\includegraphics[width=7in]{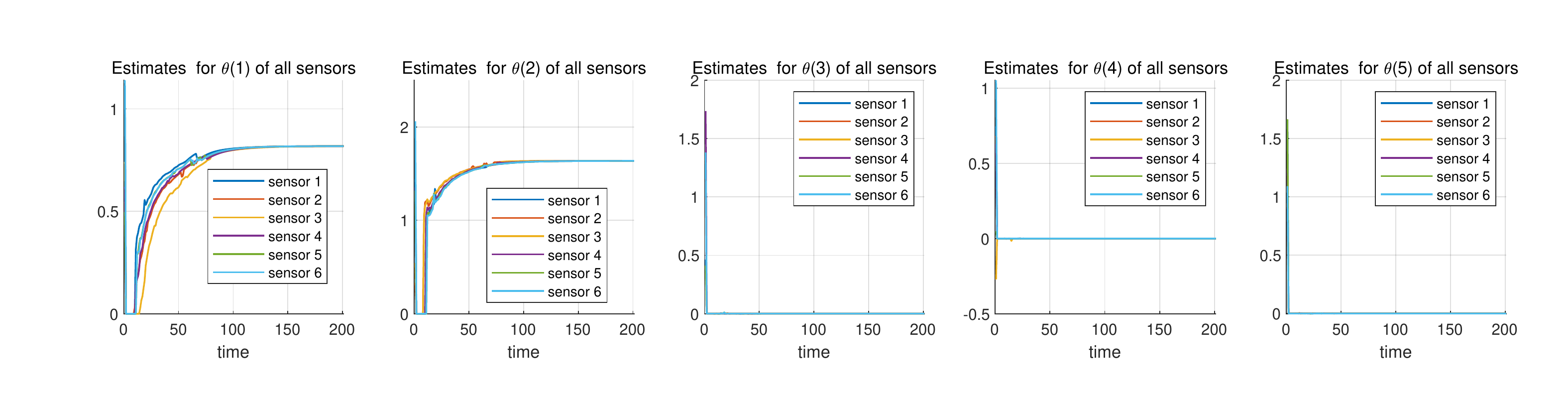}
\caption{ The estimate sequences
$\{\bm\xi_{t,i}\}^{200}_{t=0} $ of all sensors}\label{estimate}
\end{figure*}

1) We estimate the unknown  parameter $\bm\theta$ by using the non-cooperative sparse identification algorithm (i.e., the adjacency matrix is the unit matrix) and the distributed sparse identification algorithm (Algorithm \ref{algorithm2}) proposed in this paper respectively. We adopt the Matlab CVX tools
(http://cvxr.com/cvx/) to solve the convex optimization problem (\ref{sparse4}), and take the weight coefficient as $\alpha_{t,i}=( \lambda_{\min}(\bm P^{-1}_{t+1,i}))^{0.75}$. The average estimation error generated by these two algorithms is shown in Fig. \ref{compare}. We see that the estimation error generated by distributed sparse identification algorithm
converges to zero as $t$ increases, while the estimation error of the non-cooperative sparse identification algorithm does not.  The estimate sequences
$\{\bm\xi_{t,i}(1),\bm\xi_{t,i}(2),\bm\xi_{t,i}(3),\bm\xi_{t,i}(4),\bm\xi_{t,i}(5)\}^{200}_{t=0} ~(i=1,\cdots,6)$
generated by Algorithm \ref{algorithm2} are given in Fig. \ref{estimate}. We see from these figures that the estimates can converge to the true value $\bm\theta$. Therefore, the estimation
task can be fulfilled through exchanging information
between sensors even though any individual sensor can
not.

2) We estimate the unknown  parameter $\bm\theta$ by using the classical distributed LS  algorithm studied by \cite{wc2} and Algorithm \ref{algorithm2} proposed in this paper under the same network topology. Table \ref{biao1}  and Table \ref{biao2} show the estimates for $\bm\theta(3)$,  $\bm\theta(4)$, $\bm\theta(5)$ by these two algorithms at different time instants $t$. From Table \ref{biao1}  and Table \ref{biao2}, we can see that, compared
with the distributed LS algorithm in \cite{wc2},  Algorithm \ref{algorithm2} can
generate sparser and more accurate estimates for the unknown
parameters and thus give us valuable information in inferring
the zero and nonzero elements in the unknown parameters.

\begin{table*}[htb]
\caption{Estimates by the distributed LS algorithm  in \cite{wc2} and  Algorithm \ref{algorithm2} for $t=50$}\label{biao1}
\begin{center}
\begin{tabular}{lllllll}
  \hline
 ~ &\footnotesize sensor 1 &\footnotesize sensor 2 &\footnotesize sensor 3 &\footnotesize sensor 4 &\footnotesize sensor 5 &\footnotesize sensor 6 \\
  \hline
   \footnotesize Estimate for $\bm\theta(3)$  &~ & ~& ~& ~& ~& ~ \\
     \footnotesize By distributed LS & \tiny$2.5892\times10^{-4}$ &\tiny$ 1.4805\times10^{-4}$& \tiny$3.1352\times10^{-4}$&\tiny $2.7231\times10^{-4}$&  \tiny$2.9085\times10^{-4}$&\tiny $2.9085\times10^{-4}$\\
\footnotesize By Algorithm \ref{algorithm2} & \tiny$-2.8518\times10^{-6}$ &\tiny$ -6.3009\times10^{-12}$&\tiny $-8.3539\times10^{-18}$& \tiny$1.4030\times10^{-6}$& \tiny$-4.6969\times10^{-7}$& \tiny$-2.7547\times10^{-18}$ \\
  \hline
   \footnotesize Estimate for $\bm\theta(4)$  &~ & ~& ~& ~& ~& ~ \\
     \footnotesize By distributed LS& \tiny$2.7949\times10^{-4}$ &\tiny$ 2.7949\times10^{-4}$& \tiny$2.7949\times10^{-4}$&\tiny $0.0011$&  \tiny$2.7949\times10^{-4}$&\tiny $2.7949\times10^{-4}$\\
\footnotesize By Algorithm \ref{algorithm2} & \tiny$7.2376\times10^{-18}$ &\tiny$1.6087\times10^{-8}$&\tiny $-6.2511\times10^{-5}$& \tiny$1.1212\times10^{-6}$& \tiny$-3,6619\times10^{-10}$& \tiny$-7.8179\times10^{-7}$ \\
  \hline
   \footnotesize Estimate for $\bm\theta(5)$  &~ & ~& ~& ~& ~& ~ \\
     \footnotesize By distributed LS & \tiny$2.1450\times10^{-4}$ &\tiny$ 8.1487\times10^{-5}$& \tiny$1.7771\times10^{-4}$&\tiny $1.7014\times 10^{-4}$&  \tiny$4.9508\times10^{-5}$&\tiny $1.7350\times10^{-4}$\\
\footnotesize By Algorithm \ref{algorithm2} & \tiny$-2.8248\times10^{-6}$ &\tiny$1.3601\times10^{-10}$&\tiny $-3.8278\times10^{-5}$& \tiny$7.7698\times10^{-18}$& \tiny$-1.3398\times10^{-8}$& \tiny$-2.6207\times10^{-6}$ \\
  \hline
\end{tabular}
\end{center}
\end{table*}

\begin{table*}[!ht]
\caption{Estimates by the distributed LS algorithm in \cite{wc2} and Algorithm \ref{algorithm2}  for $t=100$}\label{biao2}
\begin{center}
\begin{tabular}{lllllll}
  \hline
 ~ &\footnotesize sensor 1 &\footnotesize sensor 2 &\footnotesize sensor 3 &\footnotesize sensor 4 &\footnotesize sensor 5 &\footnotesize sensor 6 \\
  \hline
   \footnotesize Estimate for $\bm\theta(3)$  &~ & ~& ~& ~& ~& ~ \\
     \footnotesize By distributed LS & \tiny$3.9929\times10^{-6}$ &\tiny$ 4.0720\times10^{-6}$& \tiny$4.5125\times10^{-6}$&\tiny $3.9929\times10^{-6}$&  \tiny$3.9929\times10^{-6}$&\tiny $4.5015\times10^{-6}$\\
\footnotesize By Algorithm \ref{algorithm2} & \tiny$ -4.1586\times10^{-13}$ &\tiny$ 	2.6792\times10^{-12}$&\tiny $1.7980\times10^{-13}$& \tiny$	-1.6160\times10^{-12}$& \tiny$8.8066\times10^{-14}$& \tiny$6.9114\times10^{-13}$ \\
  \hline
   \footnotesize Estimate for $\bm\theta(4)$  &~ & ~& ~& ~& ~& ~ \\
     \footnotesize By distributed LS & \tiny$6.4080\times10^{-6}$ &\tiny$ 3.6820\times10^{-6}$& \tiny$3.2300\times10^{-6}$&\tiny $2.5931\times10^{-6}$&  \tiny$6.4080\times10^{-6}$&\tiny $3.2300\times10^{-6}$\\
\footnotesize By Algorithm \ref{algorithm2} & \tiny$ -5.9666\times10^{-12}$ &\tiny$	-4.0833\times10^{-19}$&\tiny $-8.79535\times10^{-12}$& \tiny$-1.2865\times10^{-11}$& \tiny$	-7.3473\times10^{-12}$& \tiny$-6.2931\times10^{-12}$ \\
  \hline
   \footnotesize Estimate for $\bm\theta(5)$  &~ & ~& ~& ~& ~& ~ \\
     \footnotesize By distributed LS& \tiny$4.5652\times10^{-6}$ &\tiny$ 4.8154\times10^{-6}$& \tiny$4.6507\times10^{-6}$&\tiny $5.9311\times 10^{-6}$&  \tiny$5.5863\times10^{-6}$&\tiny $4.6507\times10^{-6}$\\
\footnotesize By Algorithm \ref{algorithm2} & \tiny$1.4196\times10^{-12}$ &\tiny$1.9062\times10^{-12}$&\tiny $-1.8454\times10^{-12}$& \tiny$3.0918\times10^{-12}$& \tiny$-2.1729\times10^{-15}$& \tiny$-1.9412\times10^{-12}$ \\
  \hline
\end{tabular}
\end{center}
\end{table*}

\section{Concluding remarks}\label{concluding}
In this paper, we first introduced a local information criterion which is formulated as a linear combination of the local estimation error with $L_1$-regularization term. By minimizing this criterion, we  proposed a distributed sparse identification algorithm to estimate an unknown parameter vector of a stochastic system. The upper bounds of the estimation error and the  averaged accumulated regrets of adaptive prediction are obtained without excitation conditions. Furthermore, we showed that under the cooperative non-persistent excitation conditions,  the set of zero elements in the unknown parameter vector can be correctly identified with a finite number of observations by properly choosing the weighting coefficient. We remark that our theoretical results are established without using such stringent conditions as independency of the regression vectors, which makes it possible to combine the distributed adaptive estimation with the distributed control. For future research, it will be interesting to consider the combination of the distributed sparse identification algorithm with the distributed control, and design a recursive distributed sparse adaptive algorithm.

\bibliographystyle{apalike}        
\bibliography{my_gd}
\appendix
	
\end{document}